\documentclass[twocolumn, twoside, journal]{IEEEtran}
\usepackage{graphicx}
\usepackage{amsmath}
\usepackage{amssymb}
\usepackage{color}

 \newcommand{\be}{\begin{equation}}
 \newcommand{\ee}{\end{equation}}
  \newcommand{\bea}{\begin{eqnarray}}
 \newcommand{\eea}{\end{eqnarray}}
 \newcommand{\ba}{\begin{array}}
\newcommand{\ea}{\end{array}}
\newcommand{\nid}{\noindent}
\newcommand{\non}{\nonumber}

\newcommand{\sss}{\scriptscriptstyle}

\IEEEoverridecommandlockouts
\renewcommand{\baselinestretch}{1.0}

\title{Waveform Design for Secure  SISO \\
 Transmissions and Multicasting \thanks{This work was supported in part by the U.S. Air
Force Office of Scientific Research (AFOSR) under Grant
FA9550-12-1-0123. This paper was presented in part at the IEEE
International Conference on Acoustics, Speech, and Signal Processing
(ICASSP), Vancouver, Canada, May 2013.}}

\begin{document}

\author{Ming~Li,~\IEEEmembership{Member,~IEEE,}
        Sandipan Kundu,
        Dimitris A. Pados,~\IEEEmembership{Member,~IEEE,}
        \\~and~Stella N. Batalama,~\IEEEmembership{Senior Member,~IEEE}
        \thanks{Ming Li, Sandipan Kundu, Dimitris A. Pados, and Stella N. Batalama are with the Department
of Electrical Engineering, State University of New York at Buffalo,
Buffalo, NY, 14260, USA (e-mail:
\{mingli,skundu,pados,batalama\}@buffalo.edu).}  }

%


\maketitle

\begin{abstract}

Wireless physical-layer security is an emerging field of research
aiming at preventing eavesdropping in an open wireless medium. In
this paper, we propose a novel waveform design approach to minimize
the likelihood that a message transmitted between trusted
single-antenna nodes is intercepted by an eavesdropper. In
particular, with knowledge first  of the eavesdropper's channel
state information (CSI), we find the optimum waveform and transmit
energy that minimize the signal-to-interference-plus-noise ratio
(SINR) at the output of the eavesdropper's maximum-SINR linear
filter, while at the same time provide the intended receiver with a
required pre-specified SINR at the output of its own max-SINR
filter. Next, if prior knowledge of the eavesdropper's CSI is
unavailable, we design a waveform that maximizes the amount of
energy available for generating disturbance to eavesdroppers, termed
artificial noise (AN), while the SINR of the intended receiver is
maintained at the pre-specified level. The extensions of the secure
waveform design problem to multiple intended receivers are also
investigated and semidefinite relaxation (SDR) -an approximation
technique based on convex optimization- is utilized to solve the
arising NP-hard design problems. Extensive simulation studies
confirm our analytical performance predictions and illustrate the
benefits of the designed waveforms on securing single-input
single-output (SISO) transmissions and multicasting.

\end{abstract}



\section{Introduction}
\label{Sec:Introduction}

\IEEEPARstart{T}{he} broadcast nature of the wireless medium makes
wireless networks ubiquitously accessible and inherently non-secure.
An eavesdropper within range of a wireless transmission may
intercept the transmitted signal while staying undetected. Commonly
used security methods rely on cryptographic (encryption) and
steganographic (covert communication) means  employed at upper
layers of the wireless network. It is still highly desirable,
however, to enhance the core security of wireless communications by
reducing the likelihood that propagating signals are intercepted by
eavesdroppers in the first place. As a result, there has been
growing interest recently in the development of physical layer
security mechanisms for the wireless link.

A classical physical-layer secrecy setting was introduced in Wyner's
seminal work \cite{Wyner 75} in the form of two single-input
single-output (SISO) channels, transmitter-to-intended-receiver and
transmitter-to-eavesdropper.  The Wyner Gaussian wiretap channel was
a first example of an information-theoretic security framework that
demonstrated the possibility of secure communications at the
physical layer. If the eavesdropper's channel is a degraded version
of the channel of the intended receiver, perfectly secure
communication between the transmitter and the intended receiver is
possible with non-zero rate. Later on,  the studies on secrecy
capacity were extended to the cases of secure communications over
SISO fading channels \cite{Barros 06}-\cite{Jeon 10}, Gaussian
broadcast channels \cite{Csiszar 78},\cite{Khisti 08}, and Gaussian
multiple access channels \cite{Tekin 06},\cite{Liang Poor 08}.
Motivated by emerging wireless communication applications with
multiple antennas, there has been recently a flurry of interesting
studies of information-theoretic secrecy capacity for multiple-input
multiple-output MIMO channels \cite{Khisti 07}-\cite{Shafiee  Ulukus
09}, single-input multiple-output (SIMO) channels \cite{Parada 05},
and  multiple-input single-output (MISO) channels \cite{Li Trappe
07}-\cite{Li 07}. Practical applications of low-density parity-check
(LDPC) codes to the wiretap channel problem were considered in
\cite{Klinc 11}-\cite{Bloch 08}.

While many works focus on information-theoretic aspects and
calculation/analysis of the achievable secrecy capacity, there is
growing interest from the signal processing perspective to provide
actual algorithmic security solutions that weaken the eavesdroppers'
intercepted signal and materialize -at least partly- the information
theoretic secrecy capacity promises. Secret transmit (and receive)
beamforming designs \cite{Khisti Wornell 10 A}-\cite{Liao 11} which
utilize the spatial degrees of freedom can enhance the physical
layer secrecy of wireless communications by crippling eavesdroppers'
interception efforts as much as possible, while simultaneously
guaranteeing a certain Quality-of-Service
(QoS)/signal-to-interference-plus-noise-ratio (SINR) at the intended
receiver. In particular, \cite{Khisti Wornell 10 A}-\cite{Khisti
Wornell 10 B} focused on exploiting knowledge of the eavesdropper's
MISO/MIMO instantaneous channel state information (CSI) to provide
secure communications. Since eavesdropper's CSI is unlikely to be
available in many scenarios, the use of artificially injected noise
(AN) was considered \cite{Goel Neg 08}-\cite{Mukherjee Swindlehurst
11}. AN-aided methods aim to generate a disturbance signal that
degrades the eavesdropper's channel but does not affect the channel
of the intended receiver, thus enabling secure communication.
AN-aided methods can certainly be adopted for the case where the
eavesdropper's instantaneous CSI is known as well. In \cite{Liao
11}, the transmit beamformer and AN spatial distribution were
jointly optimized according to the CSI of the intended receiver and
the eavesdroppers, using a semidefinite relaxation (SDR) algorithmic
approach.

%

In this present work, we consider the core problem of secure
transmissions over a multipath SISO channel where both transmitter
and intended receiver have only one antenna. Other than beamforming,
which uses the spatial degrees of freedom to weaken eavesdroppers'
receptions, we turn our attention to waveform design -another
meaningful idea in physical-layer secrecy- which can exploit the
temporal characteristics of a multipath fading channel. To the best
of our knowledge, waveform design for secure transmissions over
multipath SISO channels has not been investigated in the literature
before. Like other signal-processing-based approaches \cite{Khisti
Wornell 10 A}-\cite{Liao 11}, we will use again SINR as the
optimization metric to pursue physical-layer security. In
particular, with knowledge of eavesdropper's CSI, our objective is
to find the optimum waveform and transmit energy that minimize the
SINR at the output of the eavesdropper's maximum-SINR linear filter,
while at the same time provide the intended receiver with a
pre-specified SINR at the output of its own maximum SINR
filter\footnote{To the extend that the bit-error-rate (BER) of the
eavesdropper's receiver is monotonically decreasing in SINR,
minimization of SINR corresponds to maximization of the BER of the
eavesdropper toward the 1/2 value.}. It is also interesting to point
out that the design formulation described above is similar to
cognitive radio (CR) application problems where protecting primary
users from being interfered by secondary users \cite{Zhang
09}-\cite{Ming TWC 2011} parallels the problem of preventing
eavesdroppers from overhearing.

In the second part of this work, we study the case where no
information regarding the eavesdropper's CSI is available and
AN-aided methods are adopted in the waveform design problem. The
studies are then extended to the scenario that the transmitter is to
broadcast secure data to multiple intended receivers. We recognize
that, regretfully, the waveform design problem for secure
multicasting is non-convex NP-hard, in general. Yet, using SDR
techniques we are able to develop a realizable suboptimal solution
with excellent secure multicast system performance as demonstrated
by simulation studies included in this paper.

The rest of the paper is organized as follows. The secure SISO
transmission problem is formulated in Section \ref{Sec:System
Model}. Secure waveform designs are developed in Section
\ref{Sec:Digital Waveform Design} for one intended receiver. We then
extend the studies to the case of multiple intended receivers in
Section \ref{Sec:Multiple Legitimate Receivers}. In Section
\ref{Sec:Simulation}, simulation results illustrate our developments
and, finally, a few conclusions are drawn in Section
\ref{Sec:Conclusions}.

The following notation is used throughout the paper. Boldface
lower-case letters indicate column vectors and boldface upper-case
letters indicate matrices; $\mathbb{C}$ is the set of all complex
numbers; $()^T$ and $()^H$  denote  the transpose  and
transpose-conjugate operation, respectively;  $\mathbf{I}_L$ is the
$L\times L$ identity matrix; $\mathfrak{Re} \{\cdot\}$ denotes the
real part of a complex number; $\mathrm{sgn}\{ \cdot\}$ denotes
zero-threshold quantization; and $\mathbb{E} \{ \cdot \}$ represents
statistical expectation. $\mathbf{X}  \succ \mathbf{0}$ and
$\mathbf{X} \succeq \mathbf{0}$  state that $\mathbf{X}$ is positive
definite and positive semidefinite, respectively;
$\mathrm{Tr}\{\mathbf{X} \} $ is the trace of $\mathbf{X}$. Finally,
$| \cdot |$ and $\| \cdot \|$ are the magnitude and norm of a scalar
and vector, respectively.

\vspace{-0.0 cm}
\section{System Model}
\label{Sec:System Model} \vspace*{-0.0 cm}

\begin{center}
\begin{figure}[t!]
\begin{center}\vspace{0.3 cm}
\includegraphics[width=3.3 in]{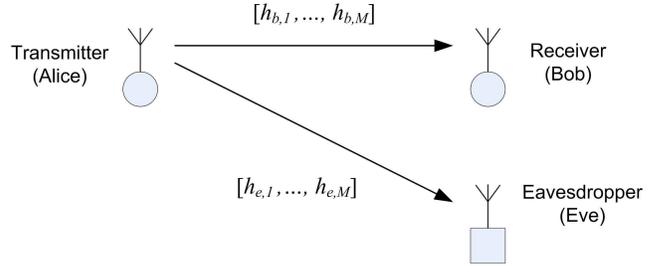}
\end{center}
\caption{SISO transmission system of a transmitter (Alice), an
intended receiver (Bob), and an eavesdropper (Eve). All received
signals exhibit multipath Rayleigh fading.} \label{fig:system model}
\end{figure}
\end{center}

We consider a wireless transmission to an  intended receiver in the
presence of an eavesdropper who is able to overhear the transmitted
signal. For convenience, we follow the common -whimsical- language
in the field and name the transmitter, intended receiver, and
eavesdropper, Alice, Bob, and Eve, respectively. A simple diagram is
shown in Fig. \ref{fig:system model} to illustrate this basic
communication scenario.

Alice will be attempting to transmit confidential messages to Bob
securely with the aid of an appropriately crafted waveform. The
transmitted signal is \be u(t) = \sum_{n=0}^{ \infty } \sqrt{E} b(n)
s (t - nT) e^{j2 \pi f_c t } \label{eq:modulation} \ee

\nid where $f_c$  is the carrier frequency, $b(n) \in \{\pm 1\}$, $n
= 1, 2, \ldots$, is the $n$th transmitted information bit, $E>0$
represents transmitted energy per bit with bit period $T$, and $s
(t)$ is the unit-energy ($ \int_0^T |s(t)|^2 dt = 1$) complex
continuous waveform  of the form
 \be s(t)  =\sum_{l=0}^{L-1}   \mathbf{s}(l)  \psi(t - l T_c)    \ee

\nid where  $\mathbf{s}(l) \in \mathbb{C}$, $l = 0, 1, \ldots , L -
1$, are to be designed/optimized and $\psi(t)$ is the continuous
pulse shaper function with duration $T_c = T/L$ assumed to be given
and fixed (for example, ideal square pulse, raised cosine, or
otherwise).

The transmitted signal is modeled to propagate to Bob and Eve over
SISO multipath Rayleigh fading channels and experience additive
white Gaussian noise (AWGN) and interference -potentially- from
other concurrent users. The combined received signal to  Bob
(subscript $b$) or Eve (subscript $e$) over individual multipath
fading channels of impulse response $h_{b/e} (t) $ is \be y_{b/e}(t)
= h_{b/e}(t)
* u (t) + z_{b/e}(t) + n_{b/e}(t) \ee

\nid where  $z_{b/e}(t)$ is other user(s) interference and
$n_{b/e}(t)$ is white Gaussian noise. After carrier demodulation and
$\psi(\cdot)$-pulse matched filtering over a presumed multipath
extended data bit period of $L_{\sss M}=L+M-1$ pulses where $M$ is
the number of resolvable multipaths, the data vector
$\mathbf{y}_b(n) \in \mathbb{C}^{L_M}$ received by Bob takes the
following general form
 \be \mathbf{y}_b(n) =
   \sqrt{E}b(n)\mathbf{H}_{b}\mathbf{s}  + \mathbf{i}_b + \textbf{z}_b
+ \mathbf{n}_b,\;\;  n=1,2, \ldots, \label{eq:Bob_1}\ee

\nid where $\mathbf{H}_{b} \in \mathbb{C}^{L_M \times L}$ is
  the multipath
 channel matrix between Alice and Bob \begin{small}
\be  \mathbf{H}_{b}  \triangleq  \left[ \begin{array}{c c c c c} h_{b,1} & 0 & \ldots & 0 & 0\\
 h_{b,2} & h_{b,1} & \ldots & 0 & 0 \\ \vdots & \vdots & \vdots & \vdots & \vdots
\\
 h_{b,M} &  h_{b,M-1}&  & 0 & 0 \\
 0 &   h_{b,M}  & & 0&  0\\
 \vdots & \vdots &  & \vdots &  \vdots\\
 0 & 0 & \ldots & h_{b,M} & h_{b,M-1}\\
  0 & 0 & \ldots & 0 & h_{b,M}
 \end{array}
\right] \label{eq:H_k} \ee \end{small}

\nid with entries $h_{b,m}  \in \mathbb{C}$, $m = 1, \ldots ,M$,
considered as complex Gaussian random variables to model fading
phenomena, $\mathbf{i}_b \in \mathbb{C}^{L_{\sss M}}$ denotes
multipath induced inter-symbol-interference (ISI), $\mathbf{z}_b \in
\mathbb{C}^{L_{\sss M}}$  represents comprehensively interference to
Bob from other  potential concurrent transmitters, and
$\mathbf{n}_b$ is a zero-mean additive white Gaussian noise (AWGN)
vector with autocorrelation matrix $\sigma^2_b \mathbf{I}_{L_{\sss
M} }$. The information bits $b(n)$ are handled as binary
equiprobable random variables that are independent within the data
stream (i.e., in $n=1, 2, \ldots$). Since the effect of ISI is,
arguably, negligible for applications of interest where the number
of resolvable multipaths $M$ is much less than the number of pulses
$L$, for mathematical and notational convenience we will not
consider the ISI terms in our theoretical developments that
follow\footnote{However, naturally, ISI will be considered and
accounted for in our simulation studies.}. Thus, Bob's received
signal in (\ref{eq:Bob_1}) is simplified/approximated by \be
\mathbf{y}_b(n) = \sqrt{E} b(n)\mathbf{H}_{b}\mathbf{s} +
\mathbf{z}_b + \mathbf{n}_b, \;\; n= 1, 2,\ldots \,.
\label{eq:Bob_2} \ee

Information bit detection at Bob is carried out optimally in
second-order statistics terms via linear maximum SINR filtering (or,
equivalently, minimum mean square error filtering) as follows \be
\widehat{b}_b (n) = \mathrm{sgn} \left\{ \mathfrak{Re} \{
\mathbf{w}^H_{maxSINR, b} \mathbf{y}_b(n)\} \right\}, \;\; n=1,2,
\ldots, \label{eq:bit detection Bob} \ee

\nid where $\mathbf{w}_{maxSINR,b} = c
\mathbf{R}_b^{-1}\mathbf{H}_{b}
 \mathbf{s} \in \mathbb{C}^{L_{ \sss M}}$, $c
> 0$, is the maximum SINR filter and $
\mathbf{R}_b \triangleq \mathbb{E}\{ (\mathbf{z}_b +\mathbf{n}_b ) (
\mathbf{z}_b +\mathbf{n}_b)^H\} = \mathbb{E}\{ \mathbf{z}_b
\mathbf{z}_b^H \}  + \sigma_b^2 \mathbf{I}_{L_{\sss M} }  \succ 0$
is the autocorrelation matrix of the combined total additive channel
disturbance. Practically, $\mathbf{R}_b$ can be estimated by
averaging signal-absent observations over $N \geq L_{\sss M}$
samples $\mathbf{y}_b(n)$
 in the absence of the signal of interest, $\mathbf{\widehat{R}}_b :=
\frac{1}{N}\sum_{n=1}^{N}[\mathbf{z}_b(n)+ \mathbf{n}_b]
[\mathbf{z}_b(n)+ \mathbf{n}_b]^H$. If interference $\mathbf{z}_b$
from other concurrent users is not present, $\mathbf{R}_b =
\sigma^2_b \mathbf{I}_{L_{\sss M}}$ and the maximum SINR filter
becomes a simple matched-filter $\mathbf{w}_{maxSINR,b} \equiv
\mathbf{w}_{MF,b} = \mathbf{H}_{b}
 \mathbf{s}$. The output SINR of
$\mathbf{w}_{maxSINR,b}$ can be calculated to be
\begin{eqnarray} \mathrm{SINR}_{b} & \triangleq & \frac{\mathbb{E}
\{ |\mathbf{w}_{maxSINR,b}^H(\sqrt{E} b \mathbf{H}_{b} \mathbf{s}
)|^2  \} }{\mathbb{E} \left\{ |\mathbf{w}_{maxSINR,b}^H
(\mathbf{z}_b + \mathbf{n}_b) |^2  \right\}
} \non \\
& = &  E\mathbf{s}^H\mathbf{H}_{b}^H \mathbf{{{R}}}_b^{-1}
\mathbf{H}_{b} \mathbf{s} =    E\mathbf{s}^H  \mathbf{Q}_b
\mathbf{s} \label{eq:SINR Bob}\end{eqnarray}

\nid where we define $\mathbf{Q}_b \triangleq  \mathbf{H}_{b}^H
\mathbf{{{R}}}_b^{-1} \mathbf{H}_{b}$, $\mathbf{Q}_b \succ 0$.

Due to the broadcast nature of the wireless medium, Eve can also
hear the signal transmitted by Alice. Without loss of generality and
for simplicity in notation, we account the multipath channels
Alice-to-Bob and Alice-to-Eve to have the same number of resolvable
paths ($M$, that is). Then, the signal vector received by Eve can be
expressed as \be \mathbf{y}_e(n) = \sqrt{E}
b(n)\mathbf{H}_{e}\mathbf{s} + \mathbf{z}_e + \mathbf{n}_e, \;\; n=
1, 2,\ldots \, , \label{eq:Eve} \ee

\nid where $\mathbf{H}_{e} \in \mathbb{C}^{L_M \times L}$ is the
Alice-to-Eve channel matrix with multipath channel coefficients
$h_{e,m} \in \mathbb{C}$, $m = 1, \ldots ,M$, $\mathbf{z}_e$ is
other-signals  interference to Eve, and $\mathbf{n}_e$ is AWGN.

We consider as a  ``worst-case'' to Alice and Bob the scenario under
which Eve has perfect knowledge of the multipath channel
coefficients $[h_{e,1}, \ldots, h_{e,M}]$ between Alice and Eve, as
well as of the waveform $\mathbf{s}$ used by Alice. Knowledge by Eve
of the waveform $\mathbf{s}$ and the Alice-to-Eve channel
coefficients $[h_{e,1}, \ldots, h_{e,M}]$ allows Eve to carry out
maximum SINR filtering eavesdropping\footnote{Knowledge by Eve of
Bob's channel $[h_{b,1}, \ldots, h_{b,M}]$ would be of no value to
passive eavesdropping, which is the only security breach considered
in this present work.}. With this information, Eve attempts to
extract/retrieve message bits via her own linear maximum SINR filter
$\mathbf{w}_{maxSINR,e}\,$, \be \hspace{-0.05 cm} \widehat{b}_e(n) =
\mathrm{sgn} \left\{ \mathfrak{Re} \{ \mathbf{w}^H_{maxSINR, e}
\mathbf{y}_e(n)\} \right\}, \, n=1,2, \ldots, \hspace{-0.05 cm}
 \label{eq: bit detection Eve} \ee

\nid where $\mathbf{w}_{maxSINR,e} = c
\mathbf{R}_e^{-1}\mathbf{H}_{e}
 \mathbf{s} \in \mathbb{C}^{L_{ \sss M}}$, $c
> 0$, and $\mathbf{R}_e \triangleq \mathbb{E}\{ \mathbf{z}_e \mathbf{z}_e^H \}  + \sigma^2_e \mathbf{I}_{L_{\sss M}
}   \succ 0$ is the autocorrelation matrix of the total additive
disturbance to Eve (which can also be sample-average estimated). The
output SINR of the filter $\mathbf{w}_{maxSINR, e}$ is given by
\begin{eqnarray} \mathrm{SINR}_{e} & \triangleq & \frac{\mathbb{E}
\{ |\mathbf{w}_{maxSINR,e}^H(\sqrt{E} b \mathbf{H}_{e} \mathbf{s}
)|^2  \} }{\mathbb{E} \left\{ |\mathbf{w}_{maxSINR,e}^H
(\mathbf{z}_e
+ \mathbf{n}_e)) |^2 \right\} \non}\\
& = &  E\mathbf{s}^H\mathbf{H}_{e}^H \mathbf{{{R}}}_e^{-1}
\mathbf{H}_{e} \mathbf{s} = E\mathbf{s}^H  \mathbf{Q}_e   \mathbf{s}
\label{eq:SINR Eve}\end{eqnarray}

\nid where we define $\mathbf{Q}_e \triangleq  \mathbf{H}_{e}^H
\mathbf{{{R}}}_e^{-1} \mathbf{H}_{e}$, $\mathbf{Q}_e \succ 0$.


From an information theoretic perspective, as long as
$\mathrm{SINR}_{b} > \mathrm{SINR}_{e}$ there exists in theory a
sequence of coding schemes in increasing block-length such that, by
adjusting the transmitting energy appropriately, \textit{only} Bob
can perfectly decode and obtain the message from Alice while Eve
fails. In a practical realistic secure wireless transmission
application, we wish that Bob can receive Alice's signal at a
minimum required SINR level that corresponds to an acceptable BER,
while Eve can only have far, far inferior SINR reception performance
with, consequently, BER near 1/2. In the next section, we attempt to
lay the foundation for such a development utilizing Alice's transmit
waveform vector $\mathbf{s}$ as a security design
parameter\footnote{Our pre-detection SINR-based development approach
is independent of symbol alphabet sets and employed detectors. For
simplicity and clarity in presentation, we consider herein binary
symbols $b(n) \in \{\pm 1 \}$ (eq. (1)) and corresponding (optimal
for Gaussian disturbance) zero-threshold detection (eqs. (7),
(10)).}.

\vspace{0.2 cm}

\section{Secure Waveform Design}
\label{Sec:Digital Waveform Design}

\subsection{Known Eavesdropper Channel}

We first consider the scenario under which Alice/Bob know Eve's
channel $\mathbf{H}_{e}$ and disturbance autocorrelation matrix
$\mathbf{R}_e$. This may be possible, for example, if the location
of Eve is known or projected/anticipated.

Our objective, in this case, is to find the transmission bit energy
$E$ and the complex-valued normalized waveform $\mathbf{s}$ used by
Alice that minimize $\mathrm{SINR}_e$ under the constraint that Bob
achieves its pre-determined SINR requirement $\gamma>0$. I.e., we
would like to identify the optimal pair \bea
 (E, \mathbf{s})^{opt} \hspace{-0.3 cm} & = & \hspace{-0.3 cm}\mathrm{arg} \underset{
E>0,\, \mathbf{s} \in \mathbb{C}^{L}}{ \mathrm{min}}  E \mathbf{s}^H
\mathbf{Q}_e \mathbf{s} \label{eq:objective 1-1}
\\
  & \mathrm{s. \;t. \;\;}  & \hspace{-0.3 cm} E \mathbf{s}^H \mathbf{Q}_b \mathbf{s}
\geq \gamma \, , \label{eq:objective 1-2} \\  & & \mathbf{s}^H
\mathbf{s} =1 \,,
  \label{eq:objective 1-3} \\ & & E \leq E_{max} \, ,   \label{eq:objective
  1-4}
\eea

\nid where $E_{max}$  denotes the maximum available/allowable bit
energy for the transmitter.

The constrained optimization problem (\ref{eq:objective
1-1})-(\ref{eq:objective 1-4}) is non-convex. It is easy to verify
that (\ref{eq:objective 1-2}) always holds with equality at an
optimal point. Therefore, for any given $\mathbf{s}$, the optimal
transmitting energy can be calculated at \be E =
\frac{\gamma}{\mathbf{s}^H \mathbf{Q}_b \mathbf{s}}. \label{eq:E}
\ee \nid  By applying (\ref{eq:E}) to (\ref{eq:objective
1-1})-(\ref{eq:objective 1-4}), the objective function can be
reformulated as having only $\mathbf{s}$ to be optimized, \bea
\mathbf{s}^{opt} \hspace{-0.3 cm} & = & \hspace{-0.3 cm}\mathrm{arg}
\; \underset{\mathbf{s} \in \mathbb{C}^{L}}{\mathrm{min}} \;
\frac{\mathbf{s}^H \mathbf{Q}_e \mathbf{s}}{\mathbf{s}^H
\mathbf{Q}_b \mathbf{s}} \label{eq:objective 2-1}\\ & \mathrm{s.
\;t. \;\;} & \hspace{-0.3 cm} \mathbf{s}^H \mathbf{Q}_b \mathbf{s}
\geq
\frac{\gamma}{E_{max}} \, ,  \label{eq:objective 2-2} \\
& & \mathbf{s}^H \mathbf{s} =1  \, .
  \label{eq:objective 2-3} \eea
%
%
%
%

\nid Now, our problem is to find a normalized waveform vector
$\mathbf{s}$ to minimize the SINR ratio (generalized Rayleigh
quotient) $\frac{\mathrm{SINR}_e}{\mathrm{SINR}_b} =
\frac{\mathbf{s}^H \mathbf{Q}_e \mathbf{s} }{\mathbf{s}^H
\mathbf{Q}_b \mathbf{s} }$ between Eve and Bob under constraint
(\ref{eq:objective 2-2}). It is clear that constraint
(\ref{eq:objective 2-2}) may be satisfied and the optimization
problem is feasible/solvable, only if the maximum eigenvalue of
$\mathbf{Q}_b$ is no less than $\gamma/E_{max}$. If we ignore
constraint (\ref{eq:objective 2-2}) for a moment, then the waveform
to minimize the SINR ratio is the familiar generalized eigenvector
solution \cite{Bie 2005} given by the following proposition.

\textit{Proposition 1}: Let $\mathbf{p}_1, \mathbf{p}_2, \ldots,
\mathbf{p}_L$ be the generalized eigenvectors of matrices
$(\mathbf{Q}_e,\mathbf{Q}_b )$ with corresponding eigenvalues
$\lambda_1 \geq \lambda_2 \geq \dots \geq \lambda_L$, i.e. $
\mathbf{Q}_e \mathbf{p}_i = \lambda_i \mathbf{Q}_b  \mathbf{p}_i$, $
i=1, \ldots, L$. The normalized waveform  to minimize the
generalized Rayleigh quotient in (\ref{eq:objective 2-1}) is the
generalized eigenvector \be \mathbf{s}=\mathbf{p}_L
\label{eq:eigen-design}\ee \nid with corresponding  smallest
eigenvalue (and attained minimum quotient/ratio) $\lambda_L$.
$\hfill \blacksquare $

The eigen-design waveform in (\ref{eq:eigen-design}) is obtained
with computational complexity  $\mathcal{O}((L+M-1)^3)$. It is
\textit{the optimal solution with Alice transmit energy} $E = \gamma
/ \mathbf{p}_L^H \mathbf{Q}_b \mathbf{p}_L$, if $\mathbf{s} =
\mathbf{p}_L$ happens to satisfy (\ref{eq:objective 2-2}), which is
a common case. If, however, (\ref{eq:objective 2-2}) is not
satisfied, we have to return to problem (\ref{eq:objective
1-1})-(\ref{eq:objective 1-4}) and examine its Karush-Kuhn-Tucker
(KKT) conditions\footnote{The strong Lagrangian duality of
(\ref{eq:objective 1-1})-(\ref{eq:objective 1-4}) was proven in
\cite{Eldar 2005}.}.  The findings are summarized in the following
proposition whose proof is provided in the Appendix.

\textit{Proposition 2}: Consider the solvable (maximum eigenvalue of
$\mathbf{Q}_b$ no less than $\gamma/E_{max}$) optimization problem
(\ref{eq:objective 2-1})-(\ref{eq:objective 2-3}) and assume that
solution (\ref{eq:eigen-design}) does not satisfy constraint
(\ref{eq:objective 2-2}). Then, the following KKT conditions are
necessary for an $\mathbf{s}$ to be  optimal \vspace*{-0.1 cm} \bea
& (\mathbf{Q}_e + \mu \mathbf{I}) \mathbf{s} = \beta
 \mathbf{Q}_b  \mathbf{s} , \; \beta > 0, \; \mu > 0 ,\; \label{eq:KKT1-1} \\
& \mathbf{s}^H \mathbf{Q}_b \mathbf{s}  =  \frac{\gamma}{E_{max}}  \; ,  \label{eq:KKT1-3}\\
&\mathbf{s}^H  \mathbf{s}  =  1  \; .  \label{eq:KKT1-4}\eea
 $\hfill \blacksquare $

While, unfortunately, we cannot have closed-form expressions for
$\mathbf{s}$ from the above KKT conditions, we can pursue an
efficient numerical solution by  bisection. We reformulate
(\ref{eq:KKT1-1}) as \be  ((1 - \tilde{\mu}) \mathbf{Q}_e +
\tilde{\mu} \mathbf{I}) \mathbf{s} =  \beta (1 - \tilde{\mu})
\mathbf{Q}_b  \mathbf{s} , \; \beta > 0, \label{eq:New_KKT1-1}\ee
where $\tilde{\mu} \triangleq \frac{\mu}{1 + \mu}$, $\tilde{\mu} \in
[0,1)$. Condition (\ref{eq:New_KKT1-1}) indicates that the optimal
$\mathbf{s}$ is a generalized eigenvector of the matrices $((1 -
\tilde{\mu}) \mathbf{Q}_e + \tilde{\mu} \mathbf{I} ,(1 -
\tilde{\mu}) \mathbf{Q}_b )$. For any given value of $\tilde{\mu}\in
[0, 1) $, let $ \mathbf{q}_{L}(\tilde{\mu})$ denote the generalized
eigenvector of $((1 - \tilde{\mu}) \mathbf{Q}_e + \tilde{\mu}
\mathbf{I} ,(1 - \tilde{\mu}) \mathbf{Q}_b )$ that has minimum
eigenvalue $\beta(\tilde\mu)$. We can easily verify that
$\mathbf{q}_{L}^H(\tilde\mu) \mathbf{Q}_b \mathbf{q}_{L}(\tilde\mu)$
is strictly monotonically increasing in $\tilde\mu \in [0, 1)$.
Based on the monotonicity and bounds on $\tilde\mu$, we solve the
KKT necessary conditions (\ref{eq:KKT1-3})-(\ref{eq:New_KKT1-1})
with bisection on $\tilde\mu$ to a value $\tilde\mu^{opt}$ such that
$| \mathbf{q}_{L}^H(\tilde\mu^{opt}) \mathbf{Q}_b
\mathbf{q}_{L}(\tilde\mu^{opt}) - \frac{\gamma}{E_{max}} |<
\epsilon$ where $\epsilon >0$ is a small positive value serving as
stopping threshold. The resulting $\tilde\mu^{opt}$,
$\beta(\tilde\mu^{opt})$, and $\mathbf{s}^{opt} =
\mathbf{q}_{L}(\tilde\mu^{opt})$ values uniquely satisfy the
necessary conditions (\ref{eq:KKT1-3})-(\ref{eq:New_KKT1-1}) and
give the globally optimal solution. While the optimization problem
can also be solved by semidefinite relaxation (SDR) \cite{Huang 10},
our proposed generalized eigen-decomposition based algorithm is
direct in nature, easy to implement (straight in the complex
domain), and faster.

\subsection{Unknown Eavesdropper Channel}

In many applications it is impractical to assume that Alice/Bob may
have (continuously updated) information about Eve's channel and
disturbance autocorrelation matrix $\mathbf{R}_e$. In this case, the
waveform design solution of the previous section cannot be adopted
due to lack of access to Eve's SINR.

By common intuition, low-power Alice-to-Bob transmission
(``whispering'') improves security by making signal interception by
Eve more difficult since Eve's SINR is proportional to the
transmitting energy. Alice, then, needs to use a waveform
$\mathbf{s}$ that  minimizes the transmitting energy while Bob
maintains a given required QoS level  \bea (E, \mathbf{s})^{opt}
\hspace{-0.2 cm} & = & \hspace{-0.2 cm}\mathrm{arg} \underset{
E>0,\, \mathbf{s} \in \mathbb{C}^{L}} {\mathrm{min}} E
\label{eq:objective 10-1}\\ & \mathrm{s. \;t. \;\;} \hspace{-0.2 cm}
&   E \mathbf{s}^H \mathbf{Q}_b \mathbf{s} \geq \gamma \, , \\  & &
\mathbf{s}^H \mathbf{s} =1 \,,
  \\ & & E \leq E_{max} \,.
\label{eq:objective 10-4}\eea

\nid  Mathematically, the optimization problem (\ref{eq:objective
10-1})-(\ref{eq:objective 10-4}) is a special case of
(\ref{eq:objective 1-1})-(\ref{eq:objective 1-4}) under
$\mathbf{Q}_e = \alpha\mathbf{I}, \alpha>0$. The optimal design to
minimize the transmit energy is summarized by the following
proposition with straightforward derivation [38, Theorem 4.2.2].

\textit{Proposition 3}: Let $\mathbf{q}_1, \mathbf{q}_2, \ldots,
\mathbf{q}_L$ be the eigenvectors of $\mathbf{Q}_b$ with
corresponding eigenvalues $\lambda_1 \geq \lambda_2 \geq \dots \geq
\lambda_L$. The waveform $\mathbf{s}$ to minimize transmitting
energy  is \be \mathbf{s}  = \mathbf{q}_1\ee \nid and the minimum
transmitting energy is \be E_{ {min}} = \gamma/ \lambda_1.
\label{eq:E_min} \ee $\hfill \blacksquare $


\nid If $E_{min} < E_{max}$,  Alice-to-Bob transmission  can be
established with waveform $\mathbf{s} = \mathbf{q}_1$ and
transmitting energy $E_{min}= \gamma/ \lambda_1$.

To further increase security by degrading Eve's SINR, we adopt an
artificial-noise (AN)-aided approach. The maximum (by the waveform
design $\mathbf{s} = \mathbf{q}_1$)  remaining transmit energy
budget $E_{AN} = E_{max} - E_{min}$ will be utilized to insert
artificially generated noise to interfere to signal reception by Eve
only. Specifically, Alice shall transmit during the $n$th symbol
period her data signal $\sqrt{E} b(n)\mathbf{s}$ along with
artificially generated noise $\mathbf{w}(n)$ of mean $\mathbb{E}\{
\mathbf{w} \} = \mathbf{0}$, autocorrelation matrix  $\mathbf{R}_w
\triangleq \mathbb{E} \{ \mathbf{w} \mathbf{w}^H \}$, and energy
$E_{AN} = \mathrm{Tr}\{ \mathbf{R}_w \} $. Bob's received signal
vector is then expressed as (compare to (6)) \be \mathbf{y}_b(n) =
\sqrt{E} b(n)\mathbf{H}_{b}\mathbf{s} + \mathbf{H}_{b}\mathbf{w}(n)
+ \mathbf{z}_b + \mathbf{n}_b, \;\; n= 1, 2,\ldots \,. \label{eq:Eve
AN} \ee

\nid With maximum SINR filtering by $\mathbf{w}_{maxSINR,b} = c
(\mathbf{R}_b + \mathbf{H}_{b} \mathbf{R}_w \mathbf{H}_{b}^H)^{-1}
\mathbf{H}_{b}\mathbf{s} $,  $c
> 0$, $\mathbf{R}_b \triangleq \mathbb{E}\{ (\mathbf{z}_b +\mathbf{n}_b )( \mathbf{z}_b +\mathbf{n}_b)^H\} $,
the output SINR with AN is maximized to \be \mathrm{SINR}_b^{AN} =
E\mathbf{s}^H \mathbf{H}_{b}^H (\mathbf{R}_b + \mathbf{H}_{b}
\mathbf{R}_w \mathbf{H}_{b}^H)^{-1} \mathbf{H}_{b} \mathbf{s}
\label{eq:SINR Bob AN}  \ee

\nid where the superscript AN is added to differentiate from Bob's
$\mathrm{SINR}_b$ in (\ref{eq:SINR Bob}) when no AN is injected by
Alice. The autocorrelation matrix $\mathbf{R}_w$ of AN must now be
designed by Alice such that Bob's SINR degradation due to AN is
zero, that is $ \mathrm{SINR}_b - \mathrm{SINR}_b^{AN} = 0$.

By Woodbury's matrix inversion lemma \cite{Meyer 00},  \bea & &
\hspace{-0.9 cm} (\mathbf{R}_b + \mathbf{H}_{b} \mathbf{R}_w
\mathbf{H}_{b}^H)^{-1} = \non
\\& & \mathbf{R}_b^{-1} - \mathbf{R}_b^{-1} \mathbf{H}_{b} \mathbf{R}_w
( \mathbf{I} + \mathbf{H}_{b}^H \mathbf{R}_b^{-1} \mathbf{H}_{b}
 \mathbf{R}_w)^{-1} \mathbf{H}_{b}^H \mathbf{R}_b^{-1} \non \eea

\nid and (\ref{eq:SINR Bob AN}) can be rewritten as \be \hspace{-0.7
cm} \mathrm{SINR}_b^{AN} =   E \mathbf{s}^H \mathbf{H}_{b}^H
\mathbf{R}_b^{-1} \mathbf{H}_{b} \mathbf{s} -  \hspace{3.1 cm} \non
\ee \be E \mathbf{s}^H \mathbf{H}_{b}^H \mathbf{R}_b^{-1}
\mathbf{H}_{b} \mathbf{R}_w ( \mathbf{I} + \mathbf{H}_{b}^H
\mathbf{R}_b^{-1} \mathbf{H}_{b}
 \mathbf{R}_w)^{-1} \mathbf{H}_{b}^H \mathbf{R}_b^{-1} \mathbf{H}_{b} \mathbf{s}  \hspace{ 0.2  cm} \label{eq:SINR AN}
 \ee

\nid where the first term is Bob's SINR without AN (see
(\ref{eq:SINR Bob})) and the second term quantifies  Bob's SINR
degradation due to AN. To make the second term (degradation) in
(\ref{eq:SINR AN}) equal to zero, it suffices to design AN with
autocorrelation matrix $\mathbf{R}_w$ such that \be \mathbf{s}^H
\mathbf{H}_{b}^H \mathbf{R}_b^{-1} \mathbf{H}_{b} \mathbf{R}_w   =
\mathbf{s}^H \mathbf{Q}_{b} \mathbf{R}_w  = \mathbf{0}^T
\label{eq:R_w equality} \ee

\nid where $\mathbf{0}$ is the $L \times 1$ all zero vector.

It is easy to see that, to achieve equality in (\ref{eq:R_w
equality}) with  waveform design $\mathbf{s} = \mathbf{q}_1$, we
should have $\mathbf{R}_w = \mathbf{W} \mathbf{\Sigma} \mathbf{W}^H
$ with $\mathbf{W} \triangleq [\mathbf{q}_2, \ldots, \mathbf{q}_L]$,
$L \geq 2$, and $\mathbf{\Sigma} \in \mathbb{R}^{(L-1) \times
(L-1)}$  a diagonal matrix with $\mathrm{Tr}\{ \mathbf{\Sigma} \} =
E_{AN}$. This means that AN $\mathbf{w}(n)$ must be chosen as a
linear combination of the $L-1$ eigenvectors $\mathbf{q}_2, \ldots,
\mathbf{q}_L $. With unknown eavesdropper's CSI, the best option
available to Alice is to isotropically/uniformly spread the
available transmit energy budget $E_{AN} = E_{max} - E_{min}$ along
the $L-1$ eigen dimensions orthogonal to $\mathbf{s} = \mathbf{q}_1$
to interfere with the eavesdroppers' receiver. Therefore, AN is
generated with the following autocorrelation matrix \be \mathbf{R}_w
= \frac{E_{max} - E_{min}}{L-1} \mathbf{W}\mathbf{W}^H.
\label{eq:R_w}   \ee

 \nid The task of weakening Eve's SINR (subject to meeting Bob's QoS
requirements) is now complete at a best effort basis by the AN
approach when neither instantaneous nor statistical CSI of Eve is
available.

\vspace{-0.1 cm}

\section{Secure Multicasting}
\label{Sec:Multiple Legitimate Receivers}

\begin{center}
\begin{figure}[t!]
\begin{center}\vspace{0.1 cm}
\includegraphics[width= 3.0 in]{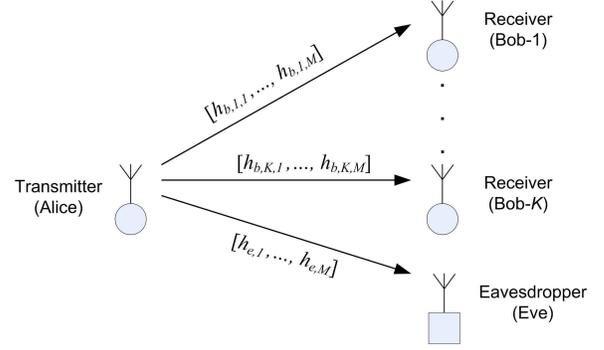}
\end{center}
\caption{Secure multicast system model of a transmitter, $K$
intended receivers, and an eavesdropper (all received signals
exhibit multipath Rayleigh fading).} \label{system
model_BC}\vspace{-0.2 cm}
\end{figure}
\end{center}

\vspace{-0.1 cm}

Multicasting is an efficient method of supporting group
communication by allowing simultaneous transmission of the same
information to multiple destinations. In the scenario of (secure)
multicasting shown in Fig. \ref{system model_BC}, Alice intends to
transmit securely the same data stream to multiple receivers ($K$
Bobs) in the presence of an eavesdropper (Eve). With $K$ Bobs to be
served,  the received signal of each Bob is denoted by \be
\mathbf{y}_{b,k}(n) = \sqrt{E} b(n)\mathbf{H}_{b,k}\mathbf{s} +
\mathbf{z}_{b,k} + \mathbf{n}_{b,k}, \non \ee \be k=1, \ldots, K,\;
n= 1, 2,\ldots \, , \non  \ee

\nid where $\mathbf{H}_{b,k} \in \mathbb{C}^{L_{\sss M} \times L}$
is the channel matrix from Alice to Bob-$k$ with  multipath channel
coefficients $h_{b,k,m} \in \mathbb{C}$, $m = 1, \ldots, M$, and
$\mathbf{z}_{b,k}$ is compound interference to Bob-$k$. Similar to
the developments in the previous section, the output SINR of
Bob-$k$'s maximum SINR filter is \be \mathrm{SINR}_{b,k} = E
\mathbf{s}^H \mathbf{Q}_{b,k} \mathbf{s} \non \ee

\nid where $\mathbf{Q}_{b,k} \triangleq \mathbf{H}_{b,k} ^H
\mathbf{R}_{b,k}^{-1} \mathbf{H}_{b,k}$, $\mathbf{R}_{b,k}
\triangleq \mathbb{E} \{ ( \mathbf{z}_{b,k} + \mathbf{n}_{b,k})
(\mathbf{z}_{b,k} + \mathbf{n}_{b,k})^{H}\} $. Eve's received signal
model is the same as in (\ref{eq:Eve}) and the output SINR of Eve's
maximum SINR filter is as in (\ref{eq:SINR Eve}).

%


If we consider sum-SINR, which is defined as the sum of the
individual SINRs of the $K$ intended receivers, as a multicast
performance metric, \be \mathrm{SINR}_{sum}   \triangleq
\sum_{k=1}^{K} \mathrm{SINR}_{b,k} = E \mathbf{s}^H
\left(\sum_{k=1}^{K} \mathbf{Q}_{b,k}\right) \mathbf{s}  =  E
\mathbf{s}^H \mathbf{\widetilde{Q}}_{b} \mathbf{s}, \non \ee

\nid  $\mathbf{\widetilde{Q}}_{b} \triangleq \sum_{k=1}^{K}
\mathbf{Q}_{b,k}$, then the presented secure waveform design problem
is similar to (\ref{eq:objective 1-1})-(\ref{eq:objective 1-4}) and
can be solved by the algorithm developed in the previous section.
Arguably, however, sum-SINR may not be an appropriate performance
measure of choice, since no form of fairness/performance assurance
among receivers can be guaranteed. Therefore, we turn our attention
to the more difficult version of the problem that involves
individual constraints by which each intended receiver has its own
SINR requirement $\gamma_k$, $k=1, \ldots, K$. We investigate secure
waveform design for the known and unknown eavesdropper channel case.

\vspace{0.2 cm}

\subsection{Known Eavesdropper Channel}

Our objective is to find the transmission bit energy $E$ and the
complex-valued normalized waveform vector $\mathbf{s}$ that minimize
$\mathrm{SINR}_e =  E \mathbf{s}^H \mathbf{Q}_{e} \mathbf{s} $ under
the constraints $\mathrm{SINR}_k = E \mathbf{s}^H \mathbf{Q}_{b,k}
\mathbf{s}  \geq \gamma_k$, $k=1, \ldots, K$, \bea  (E,
\mathbf{s})^{opt} &  \hspace{-0.2 cm} =  \hspace{-0.2 cm}&
\mathrm{arg} \underset{ E>0,\, \mathbf{s} \in \mathbb{C}^{L}}{
\mathrm{min}}  E \mathbf{s}^H \mathbf{Q}_e \mathbf{s}
\label{eq:objective 6-1}
\\ & \mathrm{s. \;t. \;\;}  \hspace{-0.2 cm} &   E \mathbf{s}^H
\mathbf{Q}_{b,k} \mathbf{s}  \geq \gamma_k, \, k=1, \ldots, K , \,
\label{eq:objective 6-2} \\  & &  \mathbf{s}^H \mathbf{s} =1 \, ,
  \label{eq:objective 6-3} \\ & & E \leq E_{max} \,.   \label{eq:objective 6-4}
\eea

\nid The optimization task of minimizing the quadratic objective
function (\ref{eq:objective 6-1}) subject to the $K>1$ constraints
in (\ref{eq:objective 6-2}) and (\ref{eq:objective 6-3}),
(\ref{eq:objective 6-4}) is, unfortunately, a non-convex NP-hard (in
$L$) optimization problem. In the following, we delve into the
details of the problem and derive a realizable suboptimum solution.

To effectively approach the problem,  we first let $\mathbf{x}
\triangleq \sqrt{E}\mathbf{s}$ denote the amplitude-including
transmitted waveform vector. Then, the optimization problem in
(\ref{eq:objective 6-1})-(\ref{eq:objective 6-4}) can be rewritten
as \bea \mathbf{x}' &  \hspace{-0.2 cm} =  \hspace{-0.2 cm} &
\mathrm{arg} \underset{\mathbf{x} \in \mathbb{C}^{L } }{
\mathrm{min}} \;
\mathbf{x}^H \mathbf{Q}_e \mathbf{x} \label{eq:objective 61-1} \\
& \mathrm{s. \;t. \;\;}   \hspace{-0.2 cm} &
\mathbf{x}^H\mathbf{Q}_{b,k} \mathbf{x} \geq \gamma_k, \, k=1,
\ldots, K , \, \label{eq:objective 61-2}
 \\ & &  \mathbf{x}^H \mathbf{x}
  \leq E_{max}\, .
   \label{eq:objective 61-3}
\eea

\nid This optimization problem is in general a non-convex
quadratically constrained quadratic program (non-convex QCQP) and
the complexity of a solver of (\ref{eq:objective
61-1})-(\ref{eq:objective 61-3}) is exponential in the dimension $L$
(NP-hard problem). To circumvent this difficulty, we first observe
that if we use the trace property of matrices, we are able to
represent the objective function in (\ref{eq:objective 61-1}) as \be
\mathbf{x}^H \mathbf{Q}_e \mathbf{x}  =
  \mathrm{Tr}\{ \mathbf{Q}_e  \mathbf{X}\} , \; \mathbf{X} \triangleq\mathbf{x}\mathbf{x}^H.\ee  \nid
Thus, with $ \mathbf{X} = \mathbf{x}\mathbf{x}^H$, the optimization
problem in (\ref{eq:objective 61-1})-(\ref{eq:objective 61-3}) takes
the new equivalent matrix form \bea  \mathbf{X}' & \hspace{-0.2 cm}
= \hspace{-0.2 cm}& \mathrm{arg} \underset{\mathbf{X} \in
\mathbb{C}^{L \times L} }{ \mathrm{min}}
\mathrm{Tr}\{ \mathbf{Q}_e  \mathbf{X}\} \label{eq:objective 7-1} \\
&\mathrm{s. \;t. \;\;}   \hspace{-0.2 cm} &  \mathrm{Tr}\{
\mathbf{Q}_{b,k} \mathbf{X}\}  \geq \gamma_k, \, k=1, \ldots, K , \,
\label{eq:objective 7-2}
 \\ & & \mathrm{Tr}\{  \mathbf{X }
 \} \leq E_{max}\, ,
   \label{eq:objective 7-3} \\
 & & \mathbf{X } \succeq 0 \; ,  \label{eq:objective 7-4} \\
& & \mathrm{rank}(\mathbf{X } ) = 1 \; . \label{eq:objective 7-5}
\eea

\nid The re-formulated design problem is, of course, still NP-hard
in general. To deal -or better say avoid- this issue, we relax/drop
the rank constraint in (\ref{eq:objective 7-5}) and solve the
simplified version by  semidefinite relaxation (SDR) \cite{Luo 10}.
The relaxed problem is a convex polynomial-complexity problem whose
optimal solution can be efficiently obtained by available
interior-point algorithms, for example the off-the-shelf solvers
\cite{Sturm 99}, \cite{Grant 09}. The worst-case computational
complexity is $\mathcal{O}((L+M-1)^{4.5} \mathrm{log}(\epsilon))$
for a given minimization solution accuracy $\epsilon > 0$.

When $\mathbf{X}'$ returned by the solver happens to be of
rank-1\footnote{By Lemma 3.1 in \cite{Huang 10}, the SDR  solution
can always be made to have rank-one when $K \leq 2$.} with
(eigenvalue, eigenvector) pair ($\lambda_1, \mathbf{a}_1$), then
$\mathbf{x}^{opt} = \sqrt{\lambda_1} \mathbf{a}_1$ and,
consequently, $E^{opt} = \lambda_1$ and $\mathbf{s}^{opt} =
\mathbf{a}_1$ for the original problem (\ref{eq:objective
6-1})-(\ref{eq:objective 6-4}). If the rank of $\mathbf{X}'$ is not
one, there is no direct path to extract $(E, \mathbf{s})^{opt}$ from
$\mathbf{X}'$  and a Gaussian randomization procedure \cite{Luo 10}
can be employed to turn the SDR solution to an approximate solution
to (\ref{eq:objective 6-1})-(\ref{eq:objective 6-4}). In particular,
we can draw now a sequence of samples $\mathbf{x}_1, \mathbf{x}_2,
\ldots , \mathbf{x}_N$ from $\mathcal{N}(\mathbf{0},\mathbf{X}')$,
i.e. Gaussian random variables with $\mathbf{0}$ mean and covariance
matrix $\mathbf{X}'$. We first apply  rescaling \be \mathbf{x}'_i =
\left( \underset{k=1, \ldots, K} {\mathrm{max}} \frac{\gamma_k} {
\mathbf{x}_i^H \mathbf{Q}_{b,k} \mathbf{x}_i } \right) \mathbf{x}_i,
\; i=1,\ldots, N, \label{eq:rescale} \ee

\nid and then test all $\mathbf{x}_i'$ for ``feasibility'' on the
constraints (\ref{eq:objective 61-2}) and (\ref{eq:objective 61-3}).
Among the feasible vectors (if any), we choose the one, say
$\mathbf{x}'^{(0)}$, with minimum $\mathbf{x}'^{H} \mathbf{Q}_e
\mathbf{x}'$ objective function value. Consequently, $E^{opt}$ and
$\mathbf{s}^{opt}$ for problem (\ref{eq:objective
6-1})-(\ref{eq:objective 6-4}) are set to $\widehat{E}^{opt} = |
\mathbf{x}'^{(0)}|^2$ and $\mathbf{\widehat{s}}^{opt} =
\mathbf{x}'^{(0)}/| \mathbf{x}'^{(0)}|$, respectively.

\subsection{Unknown Eavesdropper Channel}

For the unknown eavesdropper channel case, we pursue again the
artificial-noise (AN)-aided method. To maximize the available energy
to generate AN, we first aim at minimizing the transmitting energy
while, still, each Bob's SINR is no less than a threshold
$\gamma_k$, \bea (E, \mathbf{s} )^{opt} &  \hspace{-0.2 cm} =
\hspace{-0.2 cm} & \mathrm{arg} \underset{E>0 , \; \mathbf{s} \in
\mathbb{C}^{L} }{ \mathrm{min}} E \label{eq:objective 8-1}
\\ & \mathrm{s. \;t. \;\;}  \hspace{-0.2 cm} & E \mathbf{s}^H \mathbf{Q}_{b,k} \mathbf{s} \geq \gamma_k, \, k=1,
\ldots, K , \,  \label{eq:objective 8-2} \\  & & \mathbf{s}^H
\mathbf{s} =1 \, , \label{eq:objective 8-3} \\ & & E \leq E_{max}
\,. \label{eq:objective 8-4} \eea

\nid This NP-hard problem  can also be approximately solved by SDR
(with randomization) after reformulating (\ref{eq:objective
8-1})-(\ref{eq:objective 8-4}) in matrix form to
\begin{subequations} \non
\begin{align} \mathbf{X}' = \mathrm{arg} \underset{\mathbf{X} \in \mathbb{C}^{L \times L} }{
\mathrm{min}} & \mathrm{Tr}\{ \mathbf{X}\} \label{eq:objective 9-1}
\\ \mathrm{s. \;t. \;\;}   &  \mathrm{Tr}\{ \mathbf{Q}_{b,k}
\mathbf{X}\}  \geq \gamma_k, \, k=1, \ldots, K , \,
\label{eq:objective 9-2}
 \\ &  \mathrm{Tr}\{ \mathbf{X}
 \} \leq E_{max}\, ,
   \label{eq:objective 9-3} \\
 &  \mathbf{X } \succeq 0 \; ,  \label{eq:objective 9-4}
\end{align}
\end{subequations}

\nid where $\mathbf{X} \triangleq \mathbf{x} \mathbf{x}^H$,
$\mathbf{x} \triangleq \sqrt{E} \mathbf{s}$.  After obtaining the
optimal waveform $\mathbf{s}$ and the minimum energy $E_{min}$ via
the sole eigenvector of $\mathbf{X}'$ or an approximate solution via
(\ref{eq:rescale}) as before,  we turn our attention to the design
of AN with the residual energy $E_{max} - E_{min}$.

We recall the results from (\ref{eq:Eve AN})-(\ref{eq:SINR AN})
that, when Alice transmits AN $\mathbf{w}(n)$ along with the
information bearing signal, the output SINR of Bob-$k$'s maximum
SINR filter is \be \mathrm{SINR}_{b,k}^{AN}  = E \mathbf{s}^H
\mathbf{H}_{b,k}^H \mathbf{R}_{b,k}^{-1} \mathbf{H}_{b,k} \mathbf{s}
-  E \mathbf{s}^H \mathbf{H}_{b,k}^H \mathbf{R}_{b,k}^{-1}
\mathbf{H}_{b,k} \mathbf{R}_w \non \ee \be   ( \mathbf{I} +
\mathbf{H}_{b,k}^H
 \mathbf{R}_{b,k}^{-1}  \mathbf{H}_{b,k}\mathbf{R}_w )^{-1}
\mathbf{H}_{b,k}^H \mathbf{R}_{b,k}^{-1} \mathbf{H}_{b,k}
\mathbf{s}, \, \non \ee \be k=1, \ldots, K. \non \label{eq:SINR AN
BC} \ee

\nid To ensure that AN will not degrade the SINR of any Bob, AN
should be designed with autocorrelation matrix $\mathbf{R}_w$ such
that \be \mathbf{s}^H \mathbf{H}_{b,k}^H \mathbf{R}_{b,k}^{-1}
\mathbf{H}_{b,k} \mathbf{R}_w  = \mathbf{0}^T, \; \forall k =1,
\ldots, K. \label{eq:R_w equality BC}\ee

\nid Set $\mathbf{v}_k \triangleq   \mathbf{H}_{b,k}^H
\mathbf{R}_{b,k}^{-1} \mathbf{H}_{b,k} \mathbf{s} , k=1, \ldots, K$,
and $\mathbf{V} \triangleq [\mathbf{v}_1, \ldots, \mathbf{v}_K]$. To
achieve the equalities in (\ref{eq:R_w equality BC}), we need $L
\geq K+1$ and require that $\mathbf{R}_w  \perp \mathbf{V}$. Let
$\mathbf{u}_i$, $i=1, \ldots, L$, be left singular vectors  of
$\mathbf{V}$ with singular values $\lambda_1 \geq \lambda_2 \geq ,
\ldots, \lambda_L$. Isotropical AN should be designed with
autocorrelation matrix  \be \mathbf{R}_w = \frac{E_{max} -
E_{min}}{L-K}\mathbf{W} \mathbf{W}^H \non \ee \nid where $\mathbf{W}
\triangleq [\mathbf{u}_{K+1}, \ldots, \mathbf{u}_L]$.

\vspace*{-0.0 cm}
\section{Simulation Experiments}
\label{Sec:Simulation}

In this section, we present simulation results that show the average
SINR and bit-error-rate (BER) of Eve for various target performance
levels of Bob, lengths of  waveform $L$, and total energy
constraints. In all simulations, the channel is assumed to be
multipath fading with $M = 3$ resolvable paths with additive
interference from concurrent users and white Gaussian noise. The
multipath coefficients are taken to be independent complex Gaussian
random variables of mean zero and variance $1/M$. In each channel
realization, a number of concurrent users is randomly selected
between 5 and 10; for each concurrent user, the energy per-bit is
uniformly drawn from $[1,4]$ and a normalized waveform of length $L$
is arbitrarily generated from the zero-mean Gaussian distribution
and placed on the same carrier $f_c$ as Alice's signal. Finally, the
white Gaussian noise autocorrelation matrix at both Bob and Eve is
set at $\mathbf{I}_{L+2}$ (identity matrix of size
$L+2$)\footnote{Bob's and Eve's channel are then originally
statistically equivalent in the study.}.

\begin{center}
\begin{figure}[t!]
\hspace{-0.5cm}
\begin{center}
\includegraphics[width=3.6 in]{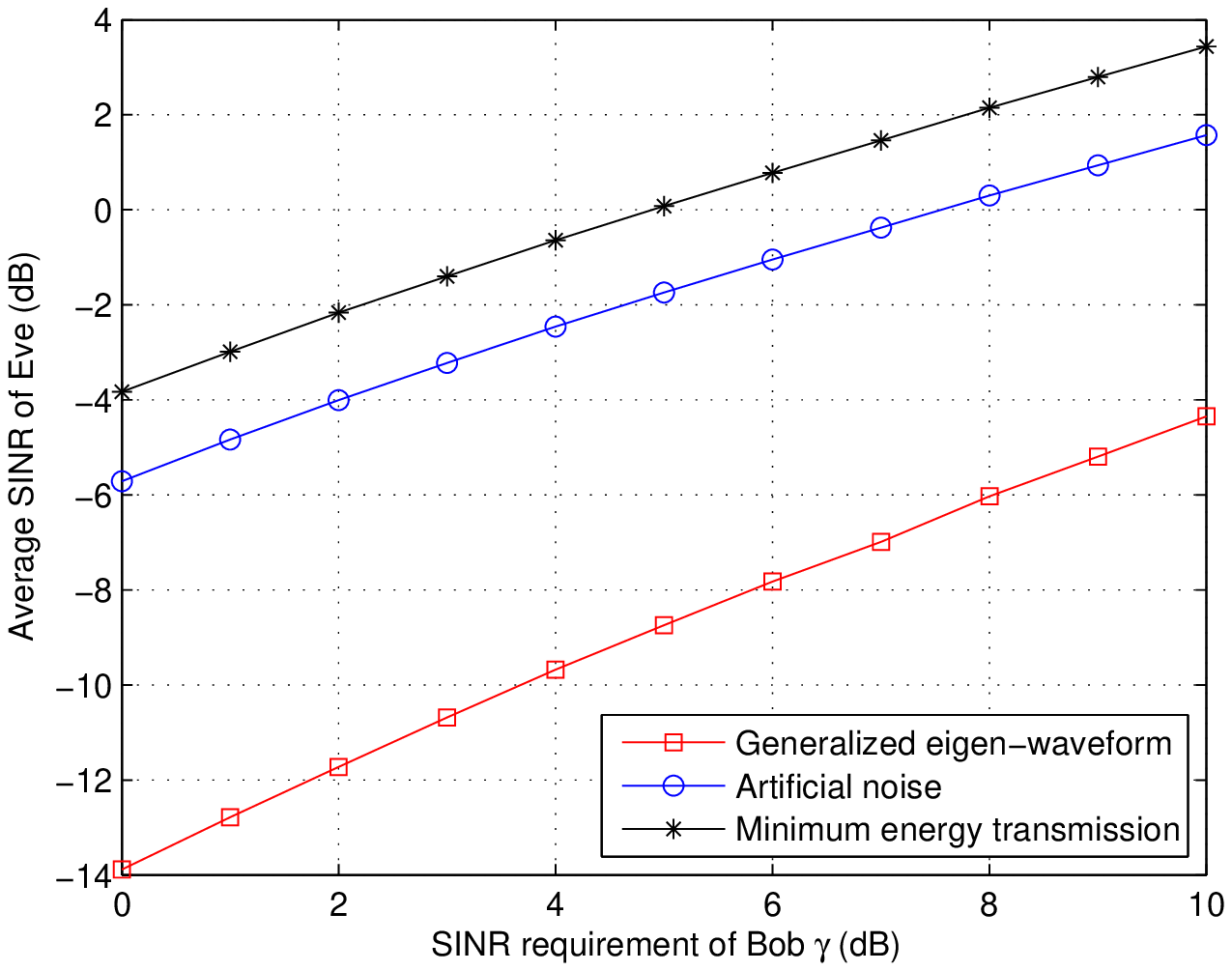}
\end{center}
\vspace{-0.3 cm} \caption{Average SINR of Eve versus SINR
requirement of Bob $\gamma$ ($E_{max} = 100$, $L=8$).}
\label{fig:SINR_SBSE_L8}
%
\vspace{-0.3 cm}
\hspace{-0.5cm}
\begin{center}
\includegraphics[width=3.6 in]{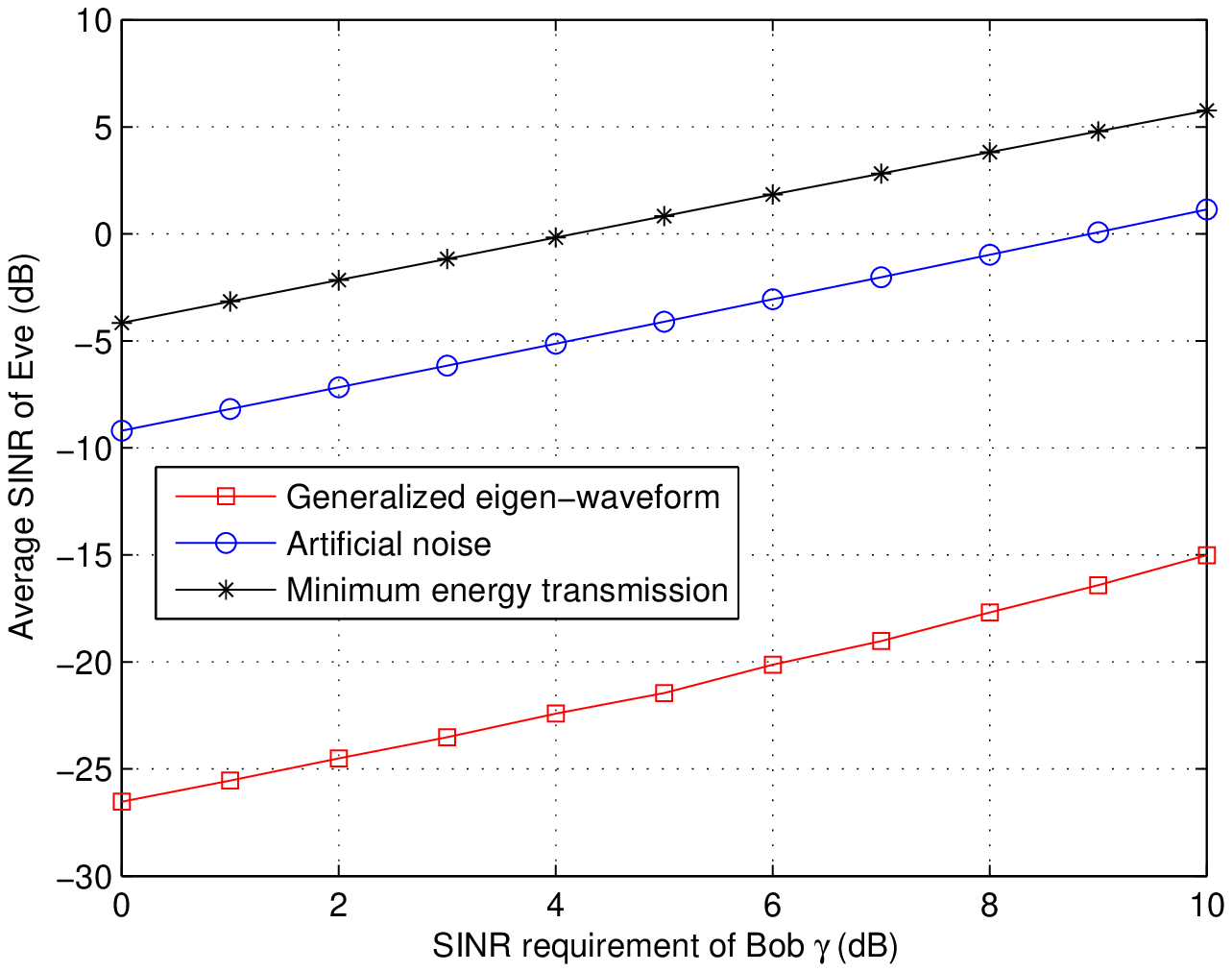}
\end{center}
\vspace{-0.5 cm} \caption{Average SINR of Eve versus SINR
requirement of Bob $\gamma$ ($E_{max} = 100$, $L=16$).}
\label{fig:SINR_SBSE_L16} \vspace{-0.0 cm}
\end{figure}
\end{center}

\vspace{-0.2 cm}

First, Alice attempts to establish a secure transmission to Bob
using a waveform of length $L=8$ in the presence of eavesdropper
Eve. The available transmit energy is assumed to be $E_{max} = 100$.
Three schemes are examined under varying assumptions about Eve's
CSI: \textit{i}) Generalized eigenwaveform of Section III.A (known
CSI); \textit{ii}) artificial noise (AN) injection of Section III.B
(no CSI); and \textit{iii}) as a reference line, minimum required
energy transmission (no CSI, no AN). The average pre-detection SINR
of Eve over $10^6$ channel realizations is plotted in Fig.
\ref{fig:SINR_SBSE_L8} as a function of Bob's pre-detection SINR
requirement $\gamma$, which is set to range from 0dB to 10dB. It can
be observed from Fig. \ref{fig:SINR_SBSE_L8} that, for the case of
known CSI, the generalized eigenwaveform design keeps the SINR of
Eve at lowest values and provides effectively secure transmission to
Bob\footnote{The transmission can be called perfectly secure if
Eve's SINR is zero or, equivalently, when her BER is 1/2. This ideal
security performance bound may not be achieved with practical system
settings, for example the short waveform length in our SISO
transmission. Nonetheless, the proposed design provides highly
effective near-optimal security, especially when the waveform length
grows to $L=16$ (Fig. \ref{fig:SINR_SBSE_L16}).}. For unknown CSI,
the AN-aided method degrades Eve's SINR by about 2dB over the no-AN
approach and maintains a significant Bob-to-Eve SINR margin of 6dB
to 8dB. In Fig. \ref{fig:SINR_SBSE_L16}, we repeat the same study
with a longer $L=16$ waveform (twice as many degrees of freedom).
Comparing Fig. \ref{fig:SINR_SBSE_L16} to Fig.
\ref{fig:SINR_SBSE_L8}, we notice the much larger SINR gains on
security even by AN alone.

\begin{center}
\begin{figure}[t!]
\hspace{-0.5 cm}
\begin{center}
\includegraphics[width=3.6 in]{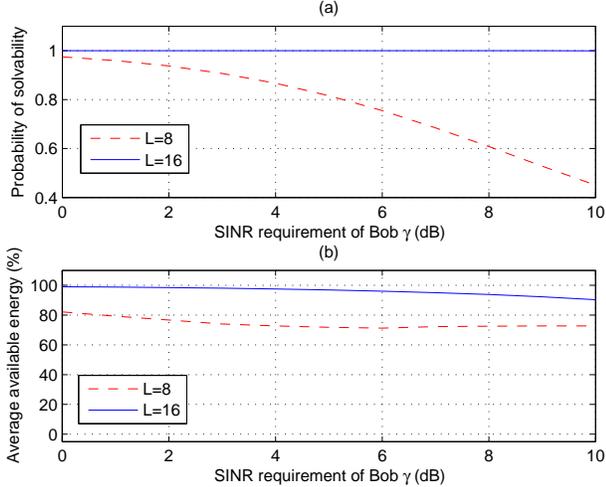}
\end{center}
\vspace{-0.2 cm} \caption{(a) Probability of solvability of the
generalized eigenwaveform optimization problem versus Bob's SINR
constraint $\gamma$ ($E_{max} = 100$); (b) average available energy
($\%$) for artificial noise creation  versus Bob's SINR constraint
$\gamma$ ($E_{max} = 100$).} \label{fig:probability} \vspace{-0.0
cm}
\end{figure}
\end{center}

In Fig. \ref{fig:probability}, we collect some useful statistics on
the experiments of Figs. \ref{fig:SINR_SBSE_L8} and
\ref{fig:SINR_SBSE_L16}. We first, Fig. \ref{fig:probability}(a),
calculate the probability (frequency of occurrence) that the
generalized eigenwaveform optimization problem in (\ref{eq:objective
1-1})-(\ref{eq:objective 1-4}) is solvable, i.e. Bob's SINR
constraint $\gamma$ can be satisfied by a waveform design for the
given $E_{max}$ value. By Fig. \ref{fig:probability}(a), the problem
is almost always solvable with $L=16$ and less likely solvable with
$L=8$ \footnote{When Alice is to transmit with an optimal
generalized eigenwaveform and no solution exists, Alice shall not
transmit to prevent eavesdropping breach.}. In Fig.
\ref{fig:probability}(b), we focus our attention on the unknown Eve
channel case and plot the average percentage of available energy to
create artificial noise. Again, $L=16$ easily supports effective
creation of AN even for large SINR requirements for Bob.

\vspace{-0.2 cm}

\begin{center}
\begin{figure}[t!]
\hspace{-0.5 cm}
\begin{center}
\includegraphics[width=3.6 in]{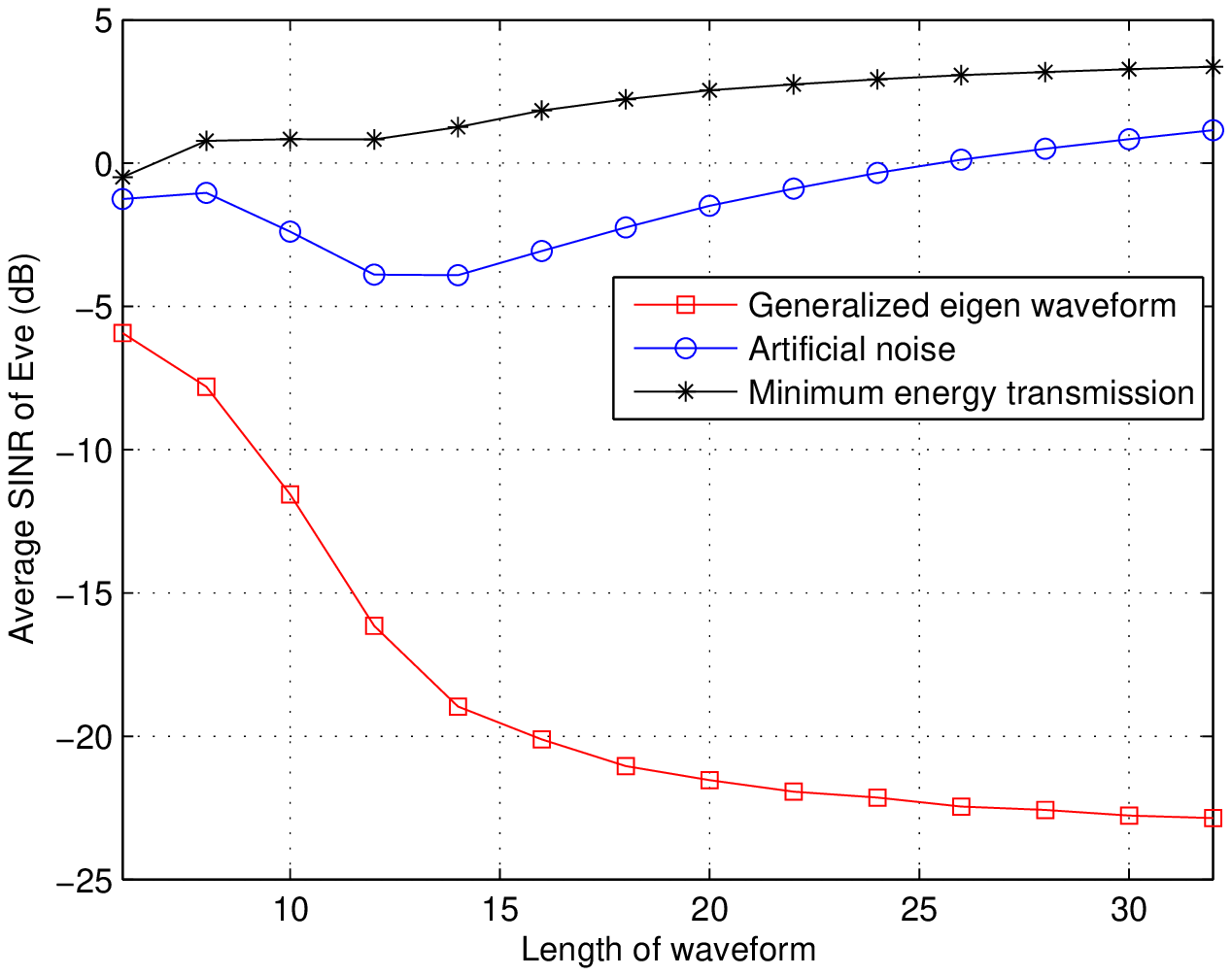}
\end{center}
\vspace{-0.3 cm} \caption{Average SINR of Eve versus length of
waveform $L$ ($E_{max} = 100$, $\gamma=6$dB).}
\label{fig:SINR_L_SBSE_SINR6}
%
\vspace{-0.3 cm}
%
\hspace{-0.5cm}
\begin{center}
\includegraphics[width=3.6 in]{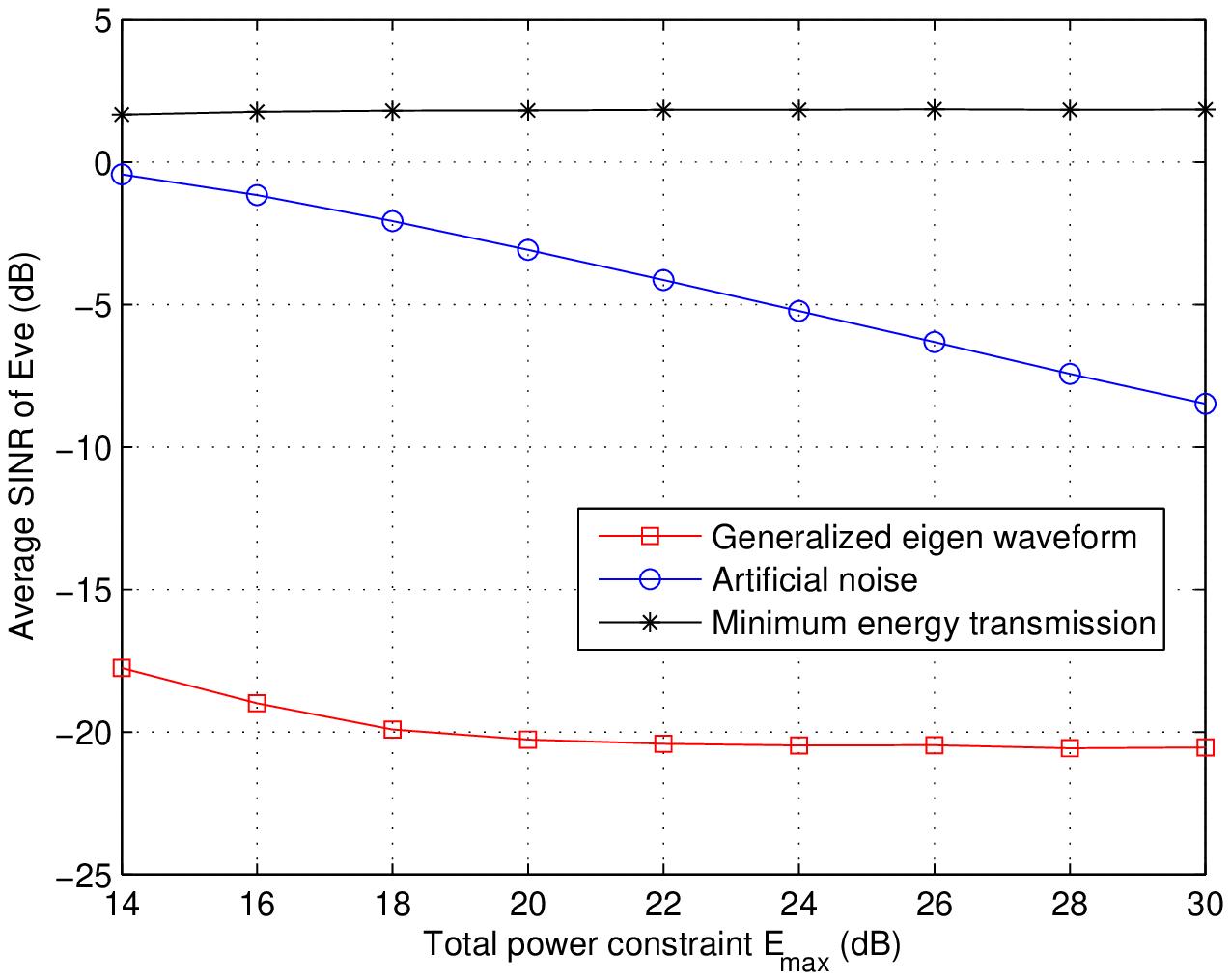}
\end{center}
\vspace{-0.3 cm} \caption{Average SINR of Eve versus total transmit
energy constraint $E_{max}$ ($L = 16$, $ \gamma =6$dB).}
\label{fig:SINR_E_SBSE_SINR6} \vspace{-0.0 cm}
\end{figure}
\end{center}

To elaborate on the relationship between security performance and
waveform length, in Fig. \ref{fig:SINR_L_SBSE_SINR6} we fix Bob's
SINR requirement at 6dB and plot the average SINR of Eve versus
waveform length $L$. While for known CSI and generalized
eigenwaveform design the average SINR of Eve continuously decreases
as $L$ increases, this is not the case for unknown CSI and AN
injection. Waveforms with longer length can reduce the transmit
energy to satisfy Bob's SINR requirement and leave more residual
energy to be used for generating AN. Ironically, while the energy of
AN can be increased by employing a longer waveform, Eve's ability to
suppress interference and noise is also enhanced\footnote{Of course,
herein, we follow the conservative approach by which Eve is supposed
to have exact knowledge of Alice's waveform.} due to the higher
space dimensions and her SINR may even increase. Therefore, waveform
length for the unknown CSI case must be selected appropriately to
balance the availability of AN energy to Alice and space dimensions
to Eve. Average SINR of Eve versus energy constraint $E_{max}$ is
shown in Fig. \ref{fig:SINR_E_SBSE_SINR6}. Obviously, for the
unknown CSI case with AN, the average SINR of Eve is decreasing with
higher energy constraint $E_{max}$, since more residual energy can
be used for generating AN.

\vspace{-0.2 cm}

\begin{center}
\begin{figure}[t!]
\hspace{-0.5cm}
\begin{center}
\includegraphics[width=3.6 in]{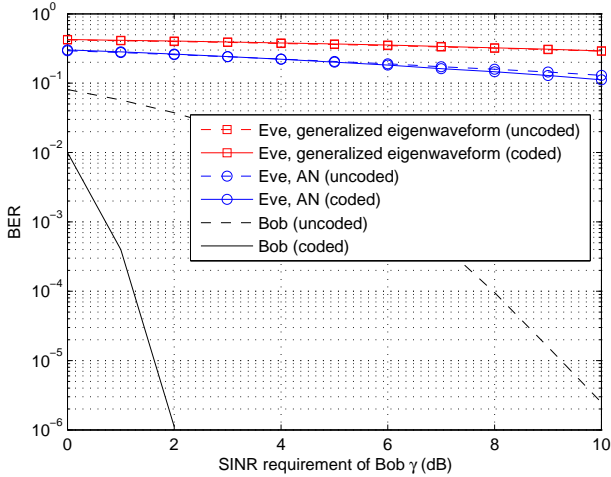}
\end{center} \vspace{-0.2 cm}
\caption{BER versus SINR requirement of Bob $\gamma$  ($E_{max} =
100$, $L=8$).} \label{fig:BER_SBSE_L8}\vspace{-0.0 cm}
\end{figure}
\end{center}

To further quantify the practical effectiveness of the proposed
transmission scheme with secure waveform design, we also evaluate
the bit-error-rate (BER) of Bob and Eve for both uncoded and coded
transmissions. An ($1024,512$) low-density parity-check (LDPC)
code\footnote{Punctured (weakened) LDPC codes were used in
\cite{Klinc 11}, \cite{Thangaraj 08} to support security. SINR-based
security optimization, as described in this present paper, places
intrinsically Bob in the ``waterfall'' region of the code while
keeping Eve ``on the top.'' Therefore, puncturing is unnecessary or
even detrimental to Bob's relative performance with respect to Eve.}
with belief-propagation decoding is adopted for the simulation
experiments. Both Bob and Eve perfectly know the coding scheme. The
BER performance curves ($L=8$) are shown in Fig.
\ref{fig:BER_SBSE_L8}. While Bob can achieve (by all practical
measures) errorless transmission with LDPC coding at SINR 2dB, Eve
has error rate barely less than $1/2$.

\vspace{-0.3 cm}

\begin{center}
\begin{figure}[t!]
\hspace{-0.5cm}
\begin{center}
\includegraphics[width=3.6 in]{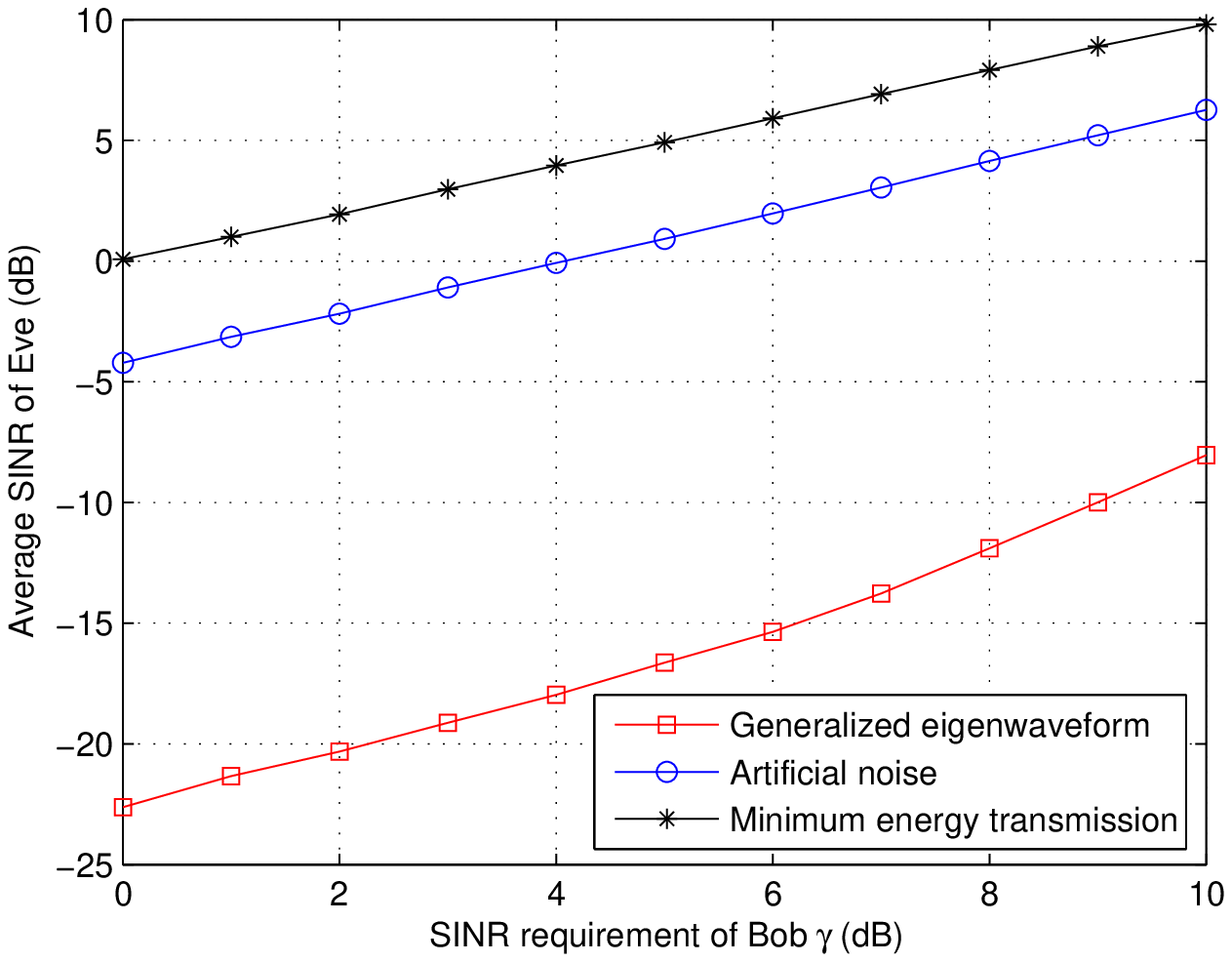}
\end{center}
\vspace{-0.2 cm} \caption{Average SINR of Eve versus SINR
requirement of Bobs $\gamma$ ($E_{max} = 100$, $L=16$,  multicast to
$K=5$ Bobs).} \label{fig:SINR_MBSE_L16}
%
\vspace{-0.3 cm}
%
\hspace{-0.5cm}
\begin{center}
\includegraphics[width=3.6 in]{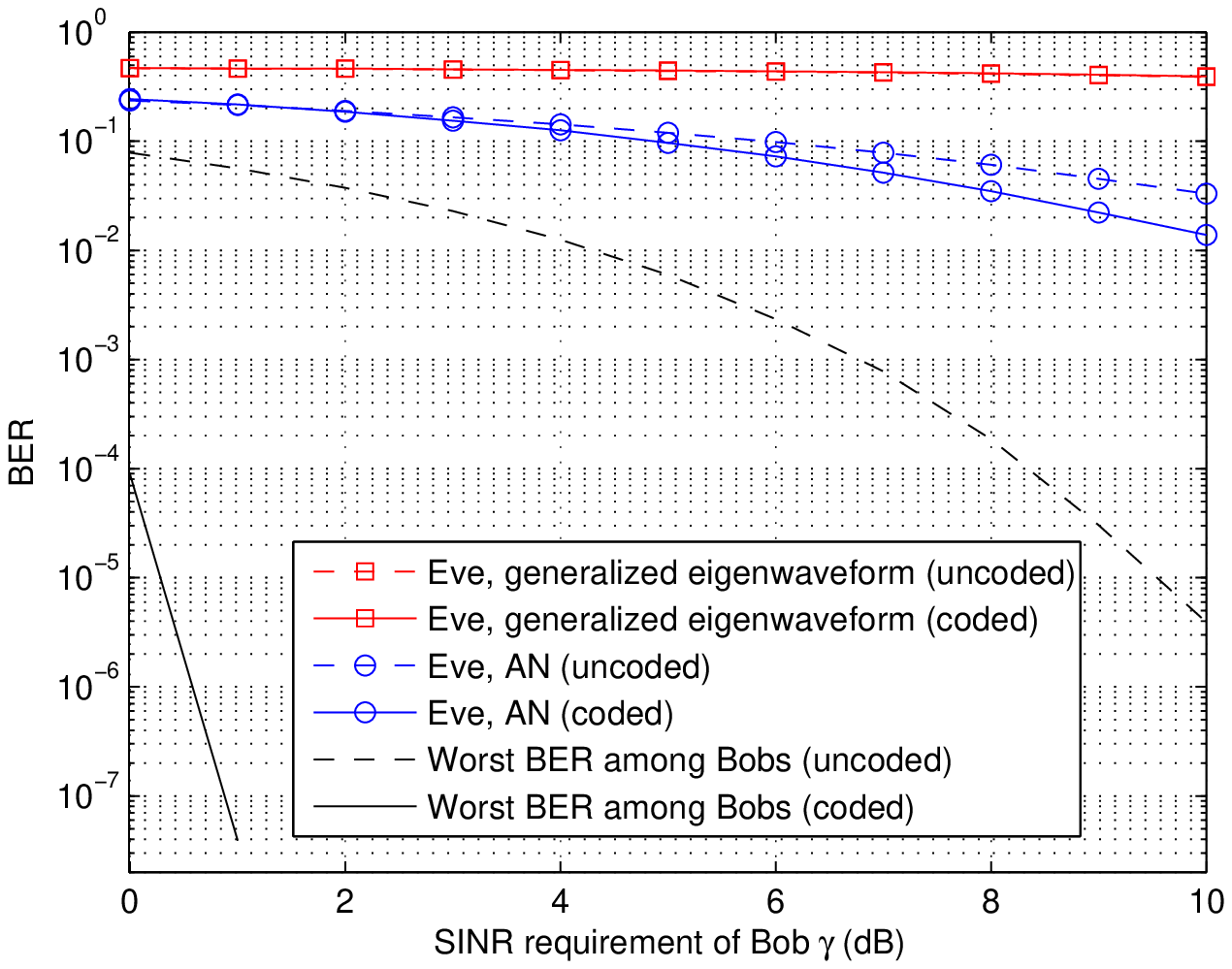}
\end{center}
\vspace{-0.4 cm} \caption{BER versus SINR requirement of Bobs
$\gamma$ ($E_{max} = 100$, $L=16$,  multicast to $K=5$ Bobs).}
\label{fig:BER_MBSE_L16} \vspace{-0.0 cm}
\end{figure}
\end{center}

Now we turn to examine the performance of secure multicasting to
multiple Bobs as described in Section \ref{Sec:Multiple Legitimate
Receivers}. Similar to the one-Bob studies, in Figs.
\ref{fig:SINR_MBSE_L16} and \ref{fig:BER_MBSE_L16} we show the
average SINR and BER, respectively, of Eve with the SDR-based
waveform designs. All $K=5$ Bobs satisfy the same SINR requirement
$\gamma_1=\ldots =\gamma_5$, which is set to vary from 0dB to 10dB.
It can be observed that the Bobs can achieve practically errorless
reception with the ($1024,512$) LDPC code when their SINR is at 2dB.
Eve's average BER is still too high (more than $10^{-1}$).

\vspace*{-0.5 cm}
\section{Conclusions}
\label{Sec:Conclusions}

We presented waveform-based approaches to secure wireless
transmissions between trusted (single-antenna) nodes in the presence
of an eavesdropper. We formulated the problem as the search for the
(transmit energy, waveform) pair that minimizes the eavesdropper's
SINR subject to the condition that the intended receiver's SINR
value is maintained at a given required SINR level (QoS determined).
A low-complexity, highly-effective eigenwaveform and transmit energy
design was proposed. We, then, extended the waveform design problem
to multiple intended receivers (secure multicasting). Regretfully,
the formulated multicasting optimization problem is non-convex and
NP-hard in the waveform dimension. Nevertheless, we employed
semi-definite relaxation to reach computationally manageable and
performance-wise appealing suboptimal solutions. Extensive
simulation experiments verified our analytical performance
predictions and illustrated the benefits of waveform optimization
for secure SISO transmission and multicasting.

As a natural next step in future work, waveform-based physical-layer
security can be combined with the existing successful beamform-based
security works in MIMO systems to carry out joint space-time
security optimization. We can harness, then, the product of space
(number of antennas) and time (waveform dimension) degrees of
freedom (DoF) to secure the link.


\appendices

\section*{Appendix - Proof of Proposition 2}

We start with the original problem (\ref{eq:objective
1-1})-(\ref{eq:objective 1-4}). We combine the function to be
optimized with the constraints and form the Lagrangian \be
\mathcal{L} = E \mathbf{s}^H \mathbf{Q}_e \mathbf{s} + \beta (
\gamma - E \mathbf{s}^H \mathbf{Q}_b \mathbf{s}) + \mu (E-E_{max}) +
\lambda (\mathbf{s}^H\mathbf{s} -1) \label{eq:Lagrangian} \ee

\nid where $\beta \geq 0$, $\mu \geq 0$, and $\lambda $ are KKT
multipliers. The KKT necessary conditions of the optimization
problem consist  of   \bea \frac{\partial \mathcal{L}}{\partial
\mathbf{s}^H} &=& E \mathbf{Q}_e \mathbf{s} - \beta E \mathbf{Q}_b
\mathbf{s} +  \lambda \mathbf{s} = \mathbf{0}, \label{eq:KKT2-1} \\
\frac{\partial \mathcal{L}}{\partial E} &=& \mathbf{s}^H
\mathbf{Q}_e \mathbf{s} - \beta \mathbf{s}^H \mathbf{Q}_b \mathbf{s}
+ \mu  =0, \label{eq:KKT2-2} \eea

\nid the complementary slackness conditions, and the primal
constraints \bea \beta (\gamma - E \mathbf{s}^H \mathbf{Q}_b
\mathbf{s}) & = & 0 \; ,  \label{eq:KKT2-3}  \\
         \mu (E-E_{max}) & = & 0 \; ,  \label{eq:KKT2-4}  \\
E \mathbf{s}^H \mathbf{Q}_b \mathbf{s} & \geq & \gamma \;  ,  \label{eq:KKT2-5}\\
\mathbf{s}^H  \mathbf{s} & = & 1  \;  ,  \label{eq:KKT2-6}\\
E & \leq &  {E_{max}} \; .  \label{eq:KKT2-7}\eea

We first examine the above KKT conditions for the cases $\beta = 0$
and $\beta > 0$, separately. If $\beta = 0$,  (\ref{eq:KKT2-2})
becomes \be \mathbf{s}^H \mathbf{Q}_e \mathbf{s} + \mu =0
\label{eq:KKT3} \ee

\nid which cannot be satisfied since $\mu \geq 0$ and
 $\mathbf{s}^H \mathbf{Q}_e \mathbf{s} > 0$. Therefore, we must have $\beta >
 0$ and \be E \mathbf{s}^H \mathbf{Q}_b \mathbf{s} = \gamma \Rightarrow
E
 = \frac{\gamma}{\mathbf{s}^H \mathbf{Q}_b \mathbf{s}}. \label{eq:KKT4}\ee

\nid We reach the equivalent problem (\ref{eq:objective
2-1})-(\ref{eq:objective 2-3}).

After applying (\ref{eq:KKT4}) to the original KKT necessary
conditions, we obtain the KKT necessary conditions for the
equivalent problem (\ref{eq:objective 2-1})-(\ref{eq:objective 2-3})
as follows \bea \mathbf{Q}_e \mathbf{s} - \beta \mathbf{Q}_b
\mathbf{s} +
\lambda (\mathbf{s}^H \mathbf{Q}_b\mathbf{s} /\gamma )   \mathbf{s}& = &\mathbf{0}, \label{eq:KKT5-1} \\
\mathbf{s}^H \mathbf{Q}_e \mathbf{s} - \beta \mathbf{s}^H
\mathbf{Q}_b \mathbf{s}
+ \mu & = & 0, \label{eq:KKT5-2}  \\
\mu (\gamma / \mathbf{s}^H \mathbf{Q}_b\mathbf{s}-E_{max}) & = & 0 , \label{eq:KKT5-3}\\
\mathbf{s}^H  \mathbf{s} & = & 1,  \label{eq:KKT5-4} \\
\mathbf{s}^H \mathbf{Q}_b \mathbf{s} & \geq &
\frac{\gamma}{E_{max}}. \label{eq:KKT5-5} \eea

\nid Left multiplying both sides of (\ref{eq:KKT5-1}) by
$\mathbf{s}^H$, we have \be \mathbf{s}^H \mathbf{Q}_e \mathbf{s} -
\beta \mathbf{s}^H \mathbf{Q}_b \mathbf{s} + \lambda (\mathbf{s}^H
\mathbf{Q}_b\mathbf{s} /\gamma )  =  0 \label{eq:KKT6-2} . \ee

\nid Combining (\ref{eq:KKT5-2}), (\ref{eq:KKT5-4}),  and
(\ref{eq:KKT6-2}), we have $\lambda (\mathbf{s}^H
\mathbf{Q}_b\mathbf{s} /\gamma )  = \mu$ and then (\ref{eq:KKT5-1})
can be rewritten as \be  (\mathbf{Q}_e + \mu \mathbf{I}) \mathbf{s}
= \beta
 \mathbf{Q}_b \mathbf{s}. \label{eq:KKT7-2} \ee

For $\mu = 0$, (\ref{eq:KKT7-2}) becomes \be \mathbf{Q}_e \mathbf{s}
=  \beta \mathbf{Q}_b \mathbf{s} \ee \nid  that implies that the
optimal waveform $\mathbf{s}$ is a generalized eigenvector of the
matrices ($ \mathbf{Q}_e, \mathbf{Q}_b$). If the solution satisfies
constraint (\ref{eq:KKT5-5}), then it is the optimal
solution; if not, then we turn to examine case $\mu > 0$. When $\mu
> 0$, to satisfy (\ref{eq:KKT5-3}), we must have $ \gamma /
\mathbf{s}^H \mathbf{Q}_b\mathbf{s}-E_{max} =0 $,  and consequently
(\ref{eq:KKT5-3}) and  (\ref{eq:KKT5-5}) together become  $
\mathbf{s}^H \mathbf{Q}_b \mathbf{s} = \frac{\gamma}{E_{max}} $.
Then, the KKT necessary conditions are (\ref{eq:KKT7-2}), together
with $\beta > 0$, $\mu > 0$, and the constraints $\mathbf{s}^H
\mathbf{R}_b \mathbf{s} = \frac{\gamma}{E_{max}} $,
$\mathbf{s}^H\mathbf{s} = 1$. The proof of Proposition 2 is
complete. $\hfill \blacksquare $

\end{document}